\documentclass{IEEEtran}

\IEEEoverridecommandlockouts
\usepackage{graphicx,psfrag,enumerate}
\usepackage{amssymb}
\usepackage{amsmath}
\usepackage{calrsfs}

\newtheorem{theorem}{Theorem}
\newtheorem{remark}{Remark}

\newtheorem{lemma}{Lemma} 
\newtheorem{corollary}{Corollary}

\usepackage{times,bm}
\usepackage{subeqnarray}
\usepackage{latexsym,psfrag}

\def\diag{\mathop{\mathrm{diag}}\nolimits} 
\def\Ker{\mathop{\mathrm{Ker}}\nolimits} 
\begin{document}

\title{Conditions for detectability in distributed
  consensus-based observer networks\thanks{This research was supported under Australian Research Council's Discovery
Projects funding scheme (Project number DP120102152).}}

\author{V. Ugrinovskii\thanks{School of Engineering and IT, University of NSW
  at the Australian Defence Force Academy, Canberra, Australia, Email:
  v.ugrinovskii@gmail.com. Part of this work was carried out during the
  author's visit to the Australian National University.}} 

\maketitle
          

\begin{abstract}
The paper discusses fundamental detectability properties
associated with the problem of distributed state estimation using networked
observers. The main result of the paper establishes connections between
detectability of the plant through measurements, observability of
the node filters through interconnections, and algebraic properties of the
underlying communication graph, to ensure the interconnected filtering
error dynamics are stabilizable via output injection. 
\end{abstract}

\section{Introduction}
An emerging trend in the area of distributed estimation is concerned with the
development of consensus-based distributed filtering algorithms to allow
each node to carry out estimation by reaching a consensus with its
neighbours. An interest in this topic is due to advantages that
distributed processing of measurements in sensor networks 
offers, over transmitting the raw data. 

A number of sufficient conditions have been proposed recently to address
the design of 
such
algorithms~\cite{SS-2009,U6,LaU1}. 
These conditions typically make use of Linear Matrix Inequalities
or matrix Riccati equations and inequalities to  
guarantee 
a suboptimal level of filter performance and/or consensus performance
between node estimators. However, the problem of establishing
feasibility of these LMI/Riccati design conditions from graph theoretical and systems theoretical viewpoints  remains an essentially open problem. 

In this paper, we consider the detectability problem for a
distributed state estimator which observes a linear plant through a network of
interconnected filters. The problem is related to a large class of distributed
estimation problems that employ interconnected observers, such as Kalman filters or $H_\infty$
filters. In particular, we are interested in the  
situation where the plant is not detectable from individual
node's measurements. For example, 
multi-vehicle Simultaneous Localization and Mapping (SLAM)
problems exhibit this feature; see the example in Section~\ref{SLAM} and~\cite{HMR-2008}. 
It was alluded in~\cite{U6} that in such situations the nodes must rely on
interconnections to ensure the state estimation problem is feasible. 
This paper
presents a rigorous analysis of this claim.    

Our chief objective is to establish conditions which guarantee
detectability of a large scale system describing observer error dynamics
in consensus based distributed estimation problems. Such
a detectability property is necessary for these estimation problems to have a solution. The
main results in this paper characterize the detectability
property of this large scale system in terms of detectability properties of
its components. Namely, we present a necessary condition for the large
scale detectability expressed in terms of the `local' detectability of the plant
through individual filter measurements, and the observability  properties
of the node filters through interconnections. Secondly, we show
that these local properties are also sufficient for the distributed
detectability property to hold if the communication graph has a spanning tree. We also extend these results to a more general 
case where the graph is weakly connected but is not spanned by a tree, and show that in this case the
problem reduces to establishing distributed detectability of certain
clusters within the system. For this, we also give necessary and sufficient
conditions. 

Our results show that in the
distributed estimation scenario, the algebraic properties of the graph
Laplacian must be complemented by observability properties of the node
filters through interconnections. This observation is in contrast
to networks of one- or two-dimensional agents, and networks consisting of
identical agents, where the ability of the system to reach consensus is
determined by the graph Laplacian matrix alone~\cite{OM-2004,RB-2005}. 

One immediate outcome of the above results concerns the design of
communications between the filter nodes. In practice, it is often 
desirable to keep transmission of information between network nodes to a
minimum, e.g., to improve the data throughput, save power, etc. The results
of this paper indicate that, as far as the detectability of the entire
system is concerned, the observability of the filters through
interconnections must be an essential design consideration. 

In regard to the role of communications, it is worthwhile to compare our
conclusions with those in~\cite{SH-2007}. The approach undertaken
in that reference is to construct interconnections to allow separation
between the agents' closed loop control dynamics and their estimator error
dynamics. In addition it achieves separation between the agents'
estimator error dynamics. This leads to the conclusion that for the
estimators to be able to converge, the system dynamics must 
be detectable from each individual node's measurements; see~\cite[Theorem
4]{SH-2007}. In contrast,  
this paper considers the case where the estimator error dynamics remain
coupled under communications. Coupling
between the error dynamics allows us to show that the 
system can be detectable, even when the plant is not detectable from
individual node's measurements.    

The paper is organized as follows. In Section~\ref{Formulation} we 
formulate the problem. The main results of the
paper are given in Section~\ref{main}. In 
Section~\ref{example}, an illustrative example is presented.
Brief conclusions are given in Section~\ref{Concl}. A conference
version of this paper was presented at the 51st IEEE CDC~\cite{U7b}.

\paragraph*{Notation} Throughout the paper, $\mathbf{R}^n$ denotes the real
Euclidean $n$-dimensional vector space. The symbol $'$ denotes the
transpose of a matrix or a 
vector. $\Ker A$ denotes the null-space of a matrix $A$. 
$\mathbf{0}_k\triangleq [0~\ldots~0]'\in \mathbf{R}^k$, 
$\mathbf{1}_k\triangleq [1~\ldots~1]'\in \mathbf{R}^k$, and
$I_k$ and $\mathbf{0}_{n\times k}$ are the identity matrix and the zero
matrix; we will omit the subscripts when this causes no ambiguity. 
The symbol $\otimes$
  denotes the Kronecker product of matrices, or the tensor product of two
  vector spaces.  $\prod_{l=1}^N\mathcal{P}_l$ will
  denote the Cartesian product of $N$ vector spaces $\mathcal{P}_1,
  \ldots,  \mathcal{P}_N$. $\dim\mathcal{X}$ is the dimension of a
  finite dimensional vector space $\mathcal{X}$. $\diag[P_1,\ldots,P_N]$
  denotes the block-diagonal matrix, whose  diagonal
  blocks are $P_1,\ldots,P_N$. 

\section{The problem formulation}\label{Formulation}

\subsection{Graph theory}
 
Consider a filter network with 
$N$ nodes and a directed graph topology $\mathbf{G} = (\mathbf{V},\mathbf{E})$;
$\mathbf{V}=\{1,2,\ldots,N\}$, $\mathbf{E}\subset \mathbf{V}\times
\mathbf{V}$ are the set of vertices and the set of edges, respectively. 
The notation $(j,i)$ will denote the edge
of the graph originating at node $j$ and ending at node $i$. In accordance
with a common convention~\cite{OM-2004}, we consider graphs without 
self-loops, i.e., $(i,i)\not\in\mathbf{E}$. However, each node is assumed
to have complete information about its filter and measurements. 

For each $i\in \mathbf{V}$, let $\mathbf{V}_i=\{j:(j,i)\in \mathbf{E}\}$
be the set of nodes supplying information to node
$i$, known as the neighbourhood of $i$.  The cardinality of
$\mathbf{V}_i$, known as the in-degree of node $i$, is denoted $p_i$; i.e.,
$p_i$ is equal to the number of incoming edges for node $i$. 
Node $i$ of a digraph is said to be reachable from node $j$ if there exists
a directed path originating at $j$ and ending at $i$. The graph is
weakly connected if any two nodes are connected by an undirected path; it is
strongly connected if its every node is reachable from any other node. 

Let $\mathbf{A}=[\mathbf{a}_{ij}]_{i,j=1}^N$ be the adjacency matrix of the
digraph $\mathbf{G}$, i.e., $\mathbf{a}_{ij}=1$ if $(j,i)\in \mathbf{E}$,
otherwise $\mathbf{a}_{ij}=0$.
Throughout the paper, $\mathcal{L}$ will denote
the $N\times N$ Laplacian matrix of the graph $\mathbf{G}$, 
$\mathcal{L}=\diag[p_1,\ldots,p_N]-\mathbf{A}$.
Since $\mathbf{G}$ has no self-loops, entries within each
  row of $\mathcal{L}$ add up to 0. Hence 0 is the eigenvalue of
$\mathcal{L}$, and $\mathbf{1}_N$ is the corresponding eigenvector. This
eigenvalue has multiplicity one if and only if the interconnection graph
has a spanning tree~\cite{RB-2005}.

\subsection{Motivating example: distributed estimation for SLAM}\label{SLAM}  

To motivate the distributed detectability problem in this paper
consider a simplified 2-D SLAM problem in which 
two robotic vehicles are required to determine the position of a static landmark
as well as the position of each other. One of the most basic models for
such SLAM system is
\begin{eqnarray}
&&\left[\begin{array}{c}\dot x_1^{(1)}\\\dot x_2^{(1)}\end{array}\right]
=\left[\begin{array}{c}\xi_x^{(1)}\\\xi_y^{(1)}\end{array}\right], \quad
\left[\begin{array}{c}\dot x_1^{(2)}\\\dot x_2^{(2)}\end{array}\right]
=\left[\begin{array}{c}\xi_x^{(2)}\\\xi_y^{(2)}\end{array}\right],
\nonumber \\
&&\left[\begin{array}{c}\dot x_1^L\\\dot x_2^L\end{array}\right]
=\left[\begin{array}{c}0 \\ 0\end{array}\right],  \label{SLAM.1}
\end{eqnarray}
where 
$x_1^{(1)},x_2^{(1)}$, $x_1^{(2)},x_2^{(2)}$, $x_1^L,x_2^L$ are
coordinates of the robots 1, 2 and the landmark, respectively;  
$\xi_{x,1},\xi_{y,1}$, $\xi_{x,2},\xi_{y,2}$ are velocity
inputs for the vehicles. The matrix
form of (\ref{SLAM.1}) is 
\begin{equation}
  \label{eq:plant}
  \dot x=Ax+B_2\xi(t), \quad x(0)=x_0,
\end{equation}
where $x=(x_1^{(1)},x_2^{(1)},x_1^{(2)},x_2^{(2)},x_1^L,x_2^L)'$ is the state
vector, and $\xi(t)=(\xi_{x,1},\xi_{y,1},\xi_{x,2},\xi_{y,2})'$. Also in this example $A=\mathbf{0}_{6\times
  6}$, $B_2=[I_4~ \mathbf{0}_{4\times
  2}]'$. Measurements used by each robot 
consist of relative robot-to-landmark measurements and measurements of
its own position (e.g., using GPS):
\begin{equation}\label{U6.yi}
y_i(t)=C_{i}x(t)+D_{i}\xi(t)+{\bar D_{i}}\xi^i(t), 
\end{equation}
where $\xi^1(t)$, $\xi^2(t)$ are measurement noises, $C_1= \left[\begin{array}{rrr} 
    -I_2 & \mathbf{0}_{2\times 2} & I_2 \\
      I_2 & \mathbf{0}_{2\times 2} & \mathbf{0}_{2\times 2}
   \end{array}\right]$,
$C_2= \left[\begin{array}{rrr}    
    \mathbf{0}_{2\times 2} & -I_2 & I_2 \\
      \mathbf{0}_{2\times 2} & I_2 & \mathbf{0}_{2\times 2}
   \end{array}\right]$, $D_{1,2}=\mathbf{0}_{4\times 4}$, $\bar D_{1,2}=I_4$.
With this notation, the SLAM problem reduces to a state estimation problem in
which each robot uses measurements (\ref{U6.yi}) to estimate the state
$x$ of the system (\ref{eq:plant}). However, it is easy to see that each of
the matrix pairs $(C_1,A)$, $(C_2,A)$ have undetectable modes, thus
rendering standard state estimation approaches infeasible.

A further analysis reveals that the undetectable subspace of $(C_1,A)$
consists of vectors $[0~0~a~b~0~0]'$ which indicates that the position of
robot 2 is not observable by robot 1. This problem will not arise if the robots
avail each other of their measurements (since the pair $([C_1'~C_2']',A)$ is
observable). Another solution is to allow robot 2 to share the estimate of
its own position with robot 1, and \emph{vice versa}. This
leads us to introduce the following distributed SLAM estimator, 
\begin{eqnarray}
    \dot{\hat x}_1&=&A\hat x_1 + L_1(y_1-C_1\hat x_1)+K_1
      (\hat z_2-H_1\hat x_1),  \nonumber \\
     \dot{\hat x}_2&=&A\hat x_2 + L_2(y_2-C_2\hat x_2)+K_2
      (\hat z_1-H_2\hat x_2). \quad \label{SLAM.filter}
\end{eqnarray} 
Here, $\hat x_1$, $\hat x_2$ denote the estimates of the vector $x$
computed by robots $1,2$, and $\hat z_1=H_2\hat x_1$, $\hat z_2=H_1\hat
x_2$ are the estimates of the robot 1 and 2 own positions, respectively, to
be shared with the neighbour;
$H_1=
\left[\mathbf{0}_{2\times 2}~ I_2~ 
  \mathbf{0}_{2\times 2}\right]$, $H_2=\left[I_2~ \mathbf{0}_{2\times 2}~
  \mathbf{0}_{2\times 2}\right]$.  

Depending on the nature of $\xi,\xi^i$ and the performance objective, the estimators in (\ref{SLAM.filter}) can be seen as Kalman
filters or $H_\infty$ filters. In both cases, the filter design is
facilitated by the fact that
the pairs $([C_i'~H_i]',A)$ are observable, and $0$ is the only state
shared by the undetectable subspace  
of $(C_i,A)$ and the observable subspace of $(H_i,A)$. 
We will show that this condition is necessary and (under additional assumptions)
sufficient for detectability of a
general class of distributed estimator networks similar to
(\ref{SLAM.filter}).
 
The interconnection matrices $H_1$,
$H_2$ given here are not unique to guarantee detectability for the SLAM filter 
(\ref{SLAM.filter}). For example, it is easy to check that using the 
`weighted disagreements' $H(\hat x_2-\hat x_1)$, $H(\hat x_1-\hat x_2)$
where $H=H_1+H_2$, instead of the `innovations'
$\hat z_2-H_1\hat x_1$, $\hat z_1-H_2\hat x_2$, will not affect the
observability and convergence properties of the filter. 
In general, we will see that the analysis of the entire observer network and its
implementation is considerably
simpler if all the agents utilize the same matrix $H$ in
their communication protocols, and the detectabilty of the network is naturally expressed in
terms of detectability properties of each network component.
However, efficient communication protocols of this form may not be so
obvious to find. The results in this paper aim at aiding in the development of
such protocols.

\subsection{The distributed detectability problem}\label{Problem}
Consider the state estimation problem for a general system of the form
(\ref{eq:plant}), 
using a network of filters connected according to the graph
$\mathbf{G}$. In (\ref{eq:plant}), $x\in\mathbf{R}^n$ is the state of the
plant, and $\xi$ denotes a  disturbance signal. 
The sensing node $i$ uses measurements of the plant given by 
(\ref{U6.yi});   
${\xi^i}$ represents the measurement uncertainty or the
measurement noise at this node, $C_i$, $D_i$, $\bar D_i$ are given matrices. Node $i$
computes its estimate of the
state $x$, denoted $\hat{x}_i\in\mathbf{R}^n$, using the filter 
 \begin{eqnarray}
    \dot{\hat x}_i=A\hat x_i + L_i(y_i(t)-C_{i}\hat x_i)+K_i\sum_{j\in
      \mathbf{V}_i}(H_i\hat x_j-H_i\hat x_i), 
  \label{UP7.C.d}
  \\
  \hat x_i(0)=0, \nonumber
\end{eqnarray}
Here $H_i$, $i=1,\ldots, N$, are given matrices.
The filter (\ref{UP7.C.d}) is a general form observer. According to (\ref{UP7.C.d}), each node computes its estimate of the plant state $x$ from its local
measurements $y_i$ and the inputs $H_i\hat x_j$ received
from its neighbours, and also communicates to the neighbours its outputs
$H_k\hat x_i$. The term 
$H_i(\hat x_j-\hat x_i)$ reflects the desire of each filter node 
to track the plant by reaching consensus with its neighbours.   
The matrices $L_i$, $K_i$ are the gain coefficients of the
filter. Depending on the  nature of disturbances and performance 
objectives, these coefficients can be determined so that the
observers (\ref{UP7.C.d}) perform as distributed Kalman filters or
distributed $H_\infty$ filters~\cite{SS-2009,U6}. 

In this paper we are not concerned with filter performance against
disturbances of a particular nature. We are interested in necessary
conditions for asymptotic convergence of every node estimator
(\ref{UP7.C.d}) to the plant in the noise-free environment, which is a natural
requirement to ensure fidelity of the estimates. Formally, it amounts to the existence of 
output injection matrices $L_i$, $K_i$, $i=1,\ldots,N$, such that the 
interconnected system consisting of the error dynamics subsystems 
 \begin{eqnarray}
    \dot{e}_i&=&\left(A - L_i
        C_i\right)e_i+K_iH_i \sum_{j\in
      \mathbf{V}_i}(e_j-e_i) \label{e.0} 
\end{eqnarray}
is globally asymptotically stable; here $e_i=x-\hat x_i$ is the local
estimation error at node $i$.  
Let $\bar A=I_N\otimes A$, $\bar C=\diag\left[C_1,\ldots,C_N\right]$, and
$\bar H=\left[\bar H_{ij}\right]_{i,j=1,\ldots,N}$ where $\bar H_{ij}=p_i
H_i$ if $j=i$, and $\bar H_{ij}=-\mathbf{a}_{ij} H_i$ if $j\neq i$,
then this requirement amounts to the detectability of 
$([\bar C',\bar H']',\bar A)$.  

From now on, we will assume identical matrices $H_i$ for all
filters (\ref{UP7.C.d}), $H_i=H$. Then $\bar H=\mathcal{L}\otimes H$. The intuition behind this assumption is drawn from the example in Section~\ref{SLAM} where the
detectability of the network was not affected when we 
replaced communication protocol matrices for both agents with judiciously
selected identical matrices. In mobile
networks with varying topology using the same matrix
$H$ may have some merits. E.g., this enables all agents to use the same
communication protocol, irrespective their location.

In the next section, we relate detectability of $([\bar C',\bar H']',\bar
A)$ with detectability properties of $(C_i,A)$,  
observability of $(H,A)$, and properties of the graph
Laplacian $\mathcal{L}$.

\section{The Main Results}\label{main}

\subsection{Geometric conditions for distributed detectability}
 
First let us recall the definition of the undetectable subspace of a matrix pair
$(G,F)$, $F\in \mathbf{R}^{n\times n}$, $G\in  
  \mathbf{R}^{m\times n}$. Let $\alpha_F(s)$ denote the minimal polynomial
    of $F$, i.e., the monic polynomial of least degree such that
$\alpha_F(F)=0$~\cite{Wonham}, factored as 
$\alpha_F(s)=\alpha_F^-(s)\alpha_F^+(s)$; the zeros of
$\alpha_F^-(s)$ and $\alpha_F^+(s)$ are in the open left and closed right
half-planes of the complex plane, respectively. Note that
$\Ker\alpha_F^-(F)\cap \Ker\alpha_F^+(F)=\{0\}$, and  $\Ker\alpha_F^-(F)+ 
\Ker\alpha_F^+(F)= \mathbf{R}^n$~\cite{Wonham}. The undetectable subspace
of $(G,F)$ is the subspace $\bigcap_{l=1}^n\Ker(GF^{l-1})\cap
\Ker\alpha_F^+(F)$~\cite{CD-1991}. When $F$ is the state matrix $A$, the
notation $O_G$ will refer to the observability matrix associated with $(G,A)$,
$
O_G=\left[\begin{array}{cccc} G' & (G A)' &
     \ldots & (G A^{n-1})'
\end{array}\right]'.
$

Consider the undetectable subspaces of $(C_i, A)$
and the unobservable subspace of $(H,A)$, which
will be denoted $\mathcal{C}_i$, $\mathcal{O}_H$. 
Furthermore, let
$\bar{\mathcal{O}}$ denote the unobservable subspace of $(\bar H, \bar A)$,  
$
\bar{\mathcal{O}}\triangleq\bigcap_{l=1}^{nN}\Ker(\bar H\bar A^{l-1}). 
$
The following general result shows that 
the large-scale system (\ref{e.0}) is detectable 
if and only if every combination of undetectable
states of the pairs $(C_i, A)$ forms an observable state of $(\bar H, \bar A)$.
\begin{lemma}\label{U7.prop1}
$([\bar C',\bar H']',\bar A)$ is detectable if and only if
\begin{equation}
\bar{\mathcal{O}}\cap \prod_{i=1}^N\mathcal{C}_i =\{0\}.
\label{undetect.C.0}
\end{equation}
\end{lemma}

The following lemma will be used in the proof of
Lemma~\ref{U7.prop1}.

\begin{lemma}\label{alpha} Recall that $\bar A=I_N\otimes A$. The following
  holds  
\begin{equation}
\Ker\alpha_{\bar A}^+(\bar A)=(\Ker \alpha_{A}^+(A))^N. 
\end{equation}
\end{lemma}

The proof of this lemma is based on the observation that $\alpha_{A}(s)$ is the
minimal polynomial for $\bar A$, and also $\alpha_{\bar
  A}^+(s)=\alpha_{A}^+(s)$. 

\emph{Proof of Lemma~\ref{U7.prop1}: }
Using Theorem~65~\cite[p.259]{CD-1991}, and the fact that
$\Ker \left[\begin{array}{c}P\\Q
\end{array}\right]=\Ker P\cap \Ker Q$,
the condition of detectability of $([\bar C',\bar H']',\bar A)$ can be
equivalently written as  
$ 
\left(\bigcap_{l=1}^{nN}\Ker(\bar C\bar
  A^{l-1})\right)\cap \Ker\alpha_{\bar A}^+(\bar A) \cap \bar{\mathcal{O}}
=\{0\}.   
$ 
Therefore to prove the lemma, we need to show that 
\begin{equation}
\left(\bigcap_{l=1}^{nN}\Ker(\bar C\bar
  A^{l-1})\right)\cap \Ker\alpha_{\bar A}^+(\bar A)=\prod_{i=1}^N
\mathcal{C}_i. 
\label{undetect.C}
\end{equation}

First, consider the set $\bigcap_{l=1}^{nN} \Ker \bar C\bar A^{l-1}$ and
take an arbitrary vector $x$ in that set, partitioned as
$x=[x_1'~\ldots~x_N']'$, $x_i\in \mathbf{R}^n$.  
Given that $\bar C$ and $\bar A$ are block diagonal,
the condition $x\in \bigcap_{l=1}^{nN} \Ker \bar C\bar A^{l-1}$ is
equivalent to $x_i\in \Ker C_iA^{l-1}$, for all $i=1,\ldots,N$ and all
$l=1,\ldots,nN$. This implies  $x_i\in \Ker O_{C_i}$ for all
$i=1,\ldots,N$. Therefore, $\bigcap_{l=1}^{nN} \Ker \bar C\bar
A^{l-1}\subseteq \prod_{i=1}^N\Ker O_{C_i}$.

Conversely, take $y_i\in \Ker O_{C_i}$. 
Using the Cayley-Hamilton theorem, this implies that  
$y=[y_1'~\ldots~y_N']'\in \Ker \bar C\bar A^{l-1}$ for all
$l=1,\ldots,nN$. This leads to the conclusion that $\prod_{i=1}^N\Ker
O_{C_i}\subseteq \bigcap_{l=1}^{nN} \Ker \bar C\bar A^{l-1}$. 
Hence,
$\bigcap_{l=1}^{nN} \Ker \bar C\bar A^{l-1}= \prod_{i=1}^N\Ker O_{C_i}$. 

To complete the proof, we now refer to Lemma~\ref{alpha}, where we showed
that $\Ker\alpha_{\bar A}^+(\bar A)=(\Ker \alpha_{A}^+(A))^N$. Since by
definition, $\mathcal{C}_i=\Ker O_{C_i}\cap \Ker \alpha_{A}^+(A)$, then
(\ref{undetect.C}) follows, as required. 
\hfill$\Box$

\begin{remark}\label{Hi}
One can see from this proof that Lemma~\ref{U7.prop1} holds in a more
general case where the matrices $H_i$ are not identical.
\end{remark}

\begin{lemma}\label{lemma.undetect.H}
Recall that $\mathcal{O}$ is the unobservable subspace
of the pair $(\bar H,\bar A)$. The following holds
\begin{equation}
\bar{\mathcal{O}}=(\Ker \mathcal{L})\otimes \mathbf{R}^n + \left(\bigcap_{l=1}^n
  \Ker (HA^{l-1})\right)^N. 
\label{undetect.H}
\end{equation}
\end{lemma}

\emph{Proof: }
First we observe that
$\bar{\mathcal{O}}=\Ker \left(\mathcal{L}\otimes O_H\right)$. 
Indeed, note that
$
(\mathcal{L}\otimes H) (I\otimes A)^{l-1}= \mathcal{L}\otimes
(HA^{l-1}). 
$ 
Hence
$
\bar{\mathcal{O}}=\Ker
\left[\begin{array}{cccc} (\mathcal{L}\otimes H)' & (\mathcal{L}\otimes (H A))' & \ldots &
    (\mathcal{L}\otimes (H A^{nN-1}))'
  \end{array}\right]'. 
$
This implies that $x=[x_1'~\ldots~x_N']\in \bar{\mathcal{O}}$ if and only if 
\begin{eqnarray}
\sum_{j\in \mathbf{V}_i} H A^{l-1}(x_i-x_j) =0, \quad
l=1,\ldots, nN.
\label{LHAx=0}
\end{eqnarray}

By the Hamilton-Caley theorem, for all $l\ge n$ one can find constants
$a_{1,l},\ldots, a_{n,l}$, such that  
$
A^lz=\sum_{\nu=1}^na_{\nu,l}(A^{\nu-1}z) \quad \forall z\in \mathbf{R}^n. 
$
Using this general identity, we establish that for all $l\ge n$, 
\begin{eqnarray*}
\sum_{j\in \mathbf{V}_i} HA^l(x_i-x_j) =
\sum_{\nu=1}^na_{\nu,l}\left(\sum_{j\in
  \mathbf{V}_i} HA^{\nu-1}(x_i-x_j) \right). 
\end{eqnarray*}
Hence, (\ref{LHAx=0}) holds for all $l=1,\ldots, nN$ if and only if it
holds for all $l=1,\ldots, n$. This proves that 
$\bar{\mathcal{O}}=\Ker\left(\mathcal{L}\otimes O_H\right)$.

Using this representation of $\bar{\mathcal{O}}$ and the fact that
$\bigcap_{l=1}^n \Ker (HA^{l-1}) =\mathcal{O}_H$, the identity 
(\ref{undetect.H}) can be re-written as
\begin{equation}
\Ker\left(\mathcal{L}\otimes O_H\right)=(\Ker \mathcal{L})\otimes \mathbf{R}^n + \left(\mathcal{O}_H\right)^N. 
\label{undetect.H.1}
\end{equation}
To prove (\ref{undetect.H.1}) we first show that
$\Ker \mathcal{L}\otimes \mathbf{R}^n + (\mathcal{O}_H)^N
\subseteq \Ker\left(\mathcal{L}\otimes O_H\right).  
$ 
It suffices to check this for elements of the subspaces 
$(\Ker \mathcal{L})\otimes \mathbf{R}^n$ and $(\mathcal{O}_H)^N$, separately.
Every element of $(\Ker \mathcal{L})\otimes \mathbf{R}^n$ is a vector of
the form $b\otimes z$, where $b\in\Ker \mathcal{L}$, and
$z\in\mathbf{R}^n$. Therefore 
$
(\mathcal{L}\otimes O_H) (b\otimes z)=\mathcal{L} b\otimes O_Hz=0. 
$
Also, choose arbitrary elements of $\mathcal{O}_H$, $h_i$, $i=1,\ldots, N$. 
Then $h=[h_1'~\ldots h_N']'\in (\mathcal{O}_H)^N$, and
\[
(\mathcal{L}\otimes O_H)h = \left[ \begin{array}{c} 
\sum_{j\in \mathbf{V}_1} O_H(h_1-h_j) \\
\vdots \\
\sum_{j\in \mathbf{V}_N} O_H(h_N-h_j)
\end{array} \right]= 0.
\]
The inclusion $\Ker \mathcal{L}\otimes \mathbf{R}^n + (\mathcal{O}_H)^N
\subseteq \Ker\left(\mathcal{L}\otimes O_H\right)  
$ 
then follows.

To prove that this inclusion 
is in fact the identity, and thus complete
the proof of the lemma, we now show that the subspaces on both sides of
the inclusion 
have the same dimension, that is
\begin{equation}
\dim ((\Ker \mathcal{L})\otimes \mathbf{R}^n +
(\mathcal{O}_H)^N) = \dim \Ker\left(\mathcal{L}\otimes O_H\right).
\label{inclusion.dim}
\end{equation}
To prove this, let $d_{\mathcal{L}}$, $d_{\mathcal{O}}$ be the dimensions
of the spaces $\Ker \mathcal{L}$, $\mathcal{O}_H$, respectively. The
following identity holds \cite{Wonham}  
\begin{eqnarray*}
\lefteqn{\dim ((\Ker \mathcal{L})\otimes \mathbf{R}^n +
(\mathcal{O}_H)^N)} &&\\
&&=nd_{\mathcal{L}}+N d_{\mathcal{O}} - \dim \left(\left(\left(\Ker
\mathcal{L}\right)\otimes \mathbf{R}^n\right) \cap (\mathcal{O}_H)^N\right). 
\end{eqnarray*}

To find the last term in the above equation, observe that a nonzero $x$
belongs to $((\Ker \mathcal{L})\otimes \mathbf{R}^n) \cap (\mathcal{O}_H)^N$
if and only if it admits the decomposition $x=[b_1z'~\ldots~b_Nz']'$ for
some $z\in \mathbf{R}^n$, $z\neq 0$ and $b=[b_1~\ldots~b_N]'\in \Ker
\mathcal{L}$, $b\neq 0$,  and also $b_iO_Hz=0$ for all $i=1,\ldots,N$. 
Since $b\neq 0$, this implies that $z\in\mathcal{O}_H$. Hence, 
$
((\Ker \mathcal{L})\otimes \mathbf{R}^n) \cap (\mathcal{O}_H)^N = (\Ker
\mathcal{L}) \otimes \mathcal{O}_H.
$
Thus, we conclude that
$
\dim ((\Ker \mathcal{L})\otimes \mathbf{R}^n +
(\mathcal{O}_H)^N)= nd_{\mathcal{L}}+(N  -
d_{\mathcal{L}}) d_{\mathcal{O}} . \quad
$

On the other hand, 
$ 
\dim \Ker(\mathcal{L}\otimes O_H)= 
nN-(N-d_{\mathcal{L}})(n-d_{\mathcal{O}})
=nd_{\mathcal{L}}+(N  - d_{\mathcal{L}}) d_{\mathcal{O}}.
$ 
Therefore, (\ref{inclusion.dim}) holds. This shows that 
the statement of the lemma holds true.
\hfill$\Box$

Our first main result, given below, presents necessary conditions for the 
detectability of the pair $([\bar C',\bar
H']',\bar A)$. 

\begin{theorem}\label{U7.prop.2}
Suppose the pair $([\bar C',\bar H']',\bar A)$ is detectable. Then, 
the following statements hold:
\begin{enumerate}[(i)]
\item
$\bigcap_{i=1}^N \mathcal{C}_i=\{0\}$;
\item 
$\mathcal{O}_H\cap \mathcal{C}_i=\{0\}$ for all $i=1,\ldots,N$;
\item
$\mathrm{rank}\, O_H \ge \max_i \dim \mathcal{C}_i$.
\end{enumerate}
\end{theorem}  

\emph{Proof: }
(i) Suppose $z\in \bigcap_{i=1}^N \mathcal{C}_i$. Then it follows from
Lemma~\ref{lemma.undetect.H} that $\mathbf{1}_N\otimes z\in (\Ker
\mathcal{L})\otimes \mathbf{R}^n\subseteq \bar{\mathcal{O}}$. Also by
  definition, $\mathbf{1}_N\otimes z\in \prod_{i=1}^N
  \mathcal{C}_i$. Hence, it follows from Lemma~\ref{U7.prop1} that
  $\mathbf{1}_N\otimes z=0$ which implies $z=0$. This proves claim (i).

(ii) Suppose $y_i\in \mathcal{O}_H\cap\mathcal{C}_i$ and consider the
vector $y=[\delta_{1i}~\delta_{2i}~\ldots~\delta_{Ni}]'\otimes y_i$, where
$\delta_{si}$ is the Kronecker symbol. By definition, $y\in
(\mathcal{O}_H)^N\subseteq \bar{\mathcal{O}}$ and $y\in \prod_{i=1}^N
  \mathcal{C}_i$. Hence, by Lemma~\ref{U7.prop1},
  $y=0$. This implies  $y_i=0$, which proves claim (ii).

(iii) From (ii), 
$
 n\ge \dim(\mathcal{O}_H+\mathcal{C}_i) 
  =\dim \mathcal{O}_H + \dim \mathcal{C}_i. 
$
Since $\mathrm{rank}\, O_H=n-\dim \mathcal{O}_H$, the claim follows.
\hfill$\Box$  

Statement (ii) of Theorem~\ref{U7.prop.2} means that for
the distributed output injection problem stated in Section~\ref{Formulation} to
have a solution, every undetectable state of $(C_i,A)$ must necessarily be an
observable state of $(H,A)$. Also, every unobservable state of $(H,A)$ must be
a detectable state of one of the pairs $(C_i,A)$.  

\subsection{Detectability over graphs spanned by trees}\label{strong} 

Our second main result shows that the conditions given in statements (i)
and (ii) of Theorem~\ref{U7.prop.2} are in fact, 
sufficient for the detectability of $([\bar C'~\bar H']',\bar A)$, provided 
the graph Laplacian matrix has a zero eigenvalue of multiplicity one. 
As is well known, this condition holds if and only if the interconnection graph
has a spanning tree~\cite{RB-2005}. The result given
in Theorem~\ref{U7.prop.4} below presents conditions on the graph
connectivity, which ensure that each node observer receives a necessary
complementary information through the interconnections.

\begin{theorem}\label{U7.prop.4}
Suppose the interconnection graph $\mathcal{G}$
has a spanning tree. If the conditions given in statements (i)
and (ii) of Theorem~\ref{U7.prop.2} hold, then the pair $([\bar
C'~\bar H']', \bar A)$  is detectable.  
\end{theorem}

\emph{Proof: }
Since $\mathcal{G}$ has a spanning tree, then the geometric multiplicity
of the zero eigenvalue of the graph Laplacian matrix $\mathcal{L}$ is equal to
1. Hence the eigenvector $\mathbf{1}_N$ is the only eigenvector (modulo scaling)
corresponding to the zero eigenvalue of $\mathcal{L}$. From this fact and
Lemma~\ref{lemma.undetect.H}, it follows that every element of 
$\bar{\mathcal{O}}$ has the form $[(z+h_1)'~\ldots~(z+h_N)']'$, where
$h_1,\ldots, h_N\in \mathcal{O}_H$, and $z$ is an arbitrary vector
$z\in\mathbf{R}^n$. 

Suppose there exists a vector of the above form which also belongs to
$\prod_{j=1}^N \mathcal{C}_j$. This implies the existence of
$z\in\mathbf{R}^n$, and $h_1,\ldots,h_N\in \mathcal{O}_H$ such that
$\forall i=1,\ldots,N,$  
\begin{eqnarray}
&&O_{C_i}z=-O_{C_i}h_i, \quad
\alpha_{A}^+(A)z=-\alpha_{A}^+(A)h_i.
\label{full.rank.system}
\end{eqnarray}
However, property (i) of Theorem~\ref{U7.prop.2} means that the matrix
$[O_{C_1}'~\ldots~O_{C_N}'~\alpha_{A}^+(A)'~\ldots~\alpha_{A}^+(A)']'$
of the system (\ref{full.rank.system})
has full row rank. Therefore, if  $z\in\mathbf{R}^n$, and
$h_1,\ldots,h_N\in \mathcal{O}_H$  satisfy (\ref{full.rank.system}), then
$z$ must be a linear combination of the vectors
$h_1\ldots,h_N$. Thus, $z\in \mathcal{O}_H$ and also $z+h_i\in
\mathcal{O}_H$ for all $i$. Using property (ii)  of
Theorem~\ref{U7.prop.2}, we conclude 
that $z+h_i=0$ for all $i=1,\ldots,N$. Hence  (\ref{undetect.C.0})
holds. According to Lemma~\ref{U7.prop1}, this means that the pair $([\bar
C'~\bar H']', \bar A)$  is detectable.   
\hfill $\Box$ 

We now specialize Theorem~\ref{U7.prop.4} to some special distributed observer
topologies commonly considered in the literature. 
The result of Corollary~\ref{spanning.tree} applies in the situation where
the root node of the graph plays the role of the leader who estimates the
plant and then passes its estimates to other nodes~\cite{LDCH-2010}. On the
contrary, 
Corollary~\ref{U7.prop.3} applies to leaderless observer networks such as
the networks connected over balanced strongly connected graphs
considered in~\cite{U6}.          

\begin{corollary}\label{U7.prop.3}
Suppose $(H,A)$ is observable.  Also, suppose  the interconnection graph
$\mathcal{L}$ has a spanning tree. If property (i) of
Theorem~\ref{U7.prop.2} holds, then the pair $([\bar C'~\bar H']', \bar A)$
is detectable.  
\end{corollary}

An immediate implication of Corollary~\ref{U7.prop.3} is that the
observability of the pair 
$(H,A)$ must be an essential design consideration when choosing a suitable matrix
$H$ for information exchange between the nodes in a leaderless network.

\begin{corollary}\label{spanning.tree}
Suppose the graph $\mathbf{G}$ has a spanning tree, with node $i$ being
the root node of the tree. Also, suppose $(C_i,A)$ is detectable at the
root node. If 
property (ii) of Theorem~\ref{U7.prop.2} holds, then the pair $([\bar
C'~\bar H']', \bar A)$  is detectable.  
\end{corollary}

\subsection{Detectability over graphs which are not spanned by a 
  tree}\label{no.tree} 

We now
restrict attention to weakly connected graphs 
 which fail
to satisfy the connectivity assumptions of
Section~\ref{strong}\footnote{If the graph is disconnected, the estimation
  problem decouples into separate estimation problems~\cite[Proposition
  1]{U6}.}.  
We show that in this case the system (\ref{e.0})
is stabilizable via output injection if and only if certain observer
clusters within the system have this property.    
 
To characterize these clusters of observers, we first discuss the relation between the structure of the interconnection graph and the
multiplicity of the zero eigenvalue of $\mathcal{L}$. The
classical result in the graph theory states that the
multiplicity of the zero eigenvalue of the Laplace matrix of an undirected
graph is equal to the number of connected components of the
graph. Recently, this result was extended to directed
graphs~\cite{AC-2005,CV-2006}. It was shown in these references that the 
multiplicity of the zero eigenvalue of $\mathcal{L}$ is equal to the number of
maximal reachable subgraphs within the graph. To present these results, some
terminology from~\cite{CV-2006} is needed.
For any vertex $j$, the reachable subgraph of $j$, $\mathbf{R}(j)$, is
defined to be the vertex subgraph containing node $j$ and all nodes
reachable from $j$.  
A vertex subgraph $\mathbf{R}$ is a reach if it is a maximal
reachable subgraph; i.e., if $\mathbf{R} = \mathbf{R}(i)$ for some $i$ and
there is no $j\neq i$ such that $\mathbf{R}(i)\subset \mathbf{R}(j)$. A
graph may consist of several reaches. For each
reach $\mathbf{R_s}$, the exclusive part of $\mathbf{R}_s$ is the
vertex subgraph $\mathbf{P}_s = \mathbf{R}_s \backslash  \cup_{r\neq
  s}\mathbf{R}_r$. The common 
part of $\mathbf{R}_s$ is the vertex subgraph
$\mathbf{Q}_s=\mathbf{R}_s\backslash \mathbf{P}_s$.  

It follows from these definitions that reaches 
have no outgoing
edges. The estimators within a reach $\mathbf{R_s}$ 
do not share
information with estimators at nodes $j\not\in \mathbf{R_s}$ but can
receive information from these nodes. On the other hand, 
the observers at nodes $i\in \mathbf{P_s}$ do not
receive information from nodes $j\not\in \mathbf{P_s}$ since by definition
$i\in \mathbf{P_s}$ is not reachable from $j\not\in \mathbf{P_s}$.

\begin{lemma}[Corollary 4.2,~\cite{CV-2006}]\label{CV.lemma}
The algebraic and geometric multiplicity of the eigenvalue 0 of
$\mathcal{L}$ is equal to the number of reaches in the graph
$\mathbf{G}$. Furthermore, $\Ker\mathcal{L}$ has a basis $b^1, \ldots,
b^k$ whose elements satisfy the conditions:
\begin{enumerate}[(i)]
\item 
$b^s_i=0$ for $i\in \mathbf{G}\backslash \mathbf{R}_s$, $s=1,\ldots,k$;
\item
$b^s_i=1$ for $i\in \mathbf{P}_s$, $s=1,\ldots,k$;
\item
$0<b^s_i<1$ for $i\in \mathbf{Q}_s$, $s=1,\ldots,k$;
\item
$\sum_{s=1}^kb^s=\mathbf{1}_N$.
\end{enumerate}
\end{lemma}
 
Theorem~3.2 in~\cite{CV-2006} shows that by permuting rows and
columns, $\mathcal{L}$ can be represented as
\begin{equation}
  \label{block.Laplacian}
  \mathcal{L}=\left[
    \begin{array}{cccc}
      \mathcal{L}_1 & \ldots & 0 & 0 \\
       0 & \ddots & 0 & 0 \\
       0 & \ldots & \mathcal{L}_k & 0 \\
       F_1 & \ldots & F_k & R 
    \end{array}
\right],
\end{equation}
where the first $k$ rows of blocks correspond to exclusive subgraphs
$\mathbf{P}_s\subset \mathbf{G}$, and the remaining rows
correspond to the vertices from $\cup_{s=1}^k\mathbf{Q}_s$. Since exclusive
subgraphs $\mathbf{P}_s$ are not reachable from the nodes outside
$\mathbf{P}_s$, each matrix $\mathcal{L}_s$, $s=1,\ldots,k$, is
a Laplacian matrix of the corresponding subgraph $\mathbf{P}_s$, and its
zero eigenvalue has multiplicity 1. Also, $R$ is shown to be invertible. 
In accordance with this partition, after the permutation the vectors $b^s$
have the form   
$
b^s=\left[\begin{array}{cccccccc}
\mathbf{0}_{l_1+\ldots+l_{s-1}}' &
    \mathbf{1}_{l_s}' & 
\mathbf{0}_{l_{s+1}+\ldots+l_k}' &
    (f^s)' 
  \end{array}\right]',
$
where $l_q=\dim\mathcal{L}_q$ is the cardinality of the vertex set of
$\mathbf{P}_q$. Also, $f^s=-R^{-1}F_s{\mathbf{1}_{l_s}}\in\mathbf{R}^r$,
$r$ being the cardinality of the vertex set of $\cup_{s=1}^k
\mathbf{Q}_s$. From Lemma~\ref{CV.lemma}, the vector $f^s$ can be further
partitioned $f^s=[(f^s_1)'~\ldots~(f^s_k)']'$, where $f^s_q$ designates the
component corresponding to the nodes of $\mathbf{Q}_q$ after the permutation.
Therefore, $f^s_q=0$ for $q\neq s$, and all the entries of $f_s^s=[f^s_{s,1}~
\ldots~f^s_{s,r_s}]' $ corresponding to the vertices in $\mathbf{Q}_s$
satisfy $0<f^s_{s,i}<1$; $r_s$ denotes the cardinality of the vertex set of
$\mathbf{Q}_s$.

\begin{theorem}\label{detect.theorem}
Suppose the pair $([\bar C',\bar H']',\bar A)$ is detectable. Then, for
every reach $\mathbf{R}_s\subset\mathbf{G}$:
\begin{enumerate}[(i)]
\item
$\bigcap_{i\in\mathbf{R}_s} \mathcal{C}_i=\{0\}$; 
\item 
$\mathcal{O}_H\cap \mathcal{C}_i=\{0\}$ for all $i\in\mathbf{R}_s$.
\end{enumerate}
\end{theorem}

\emph{Proof: }
Statement (ii) follows from Theorem~\ref{U7.prop.2}. 
Now suppose that there exists a reach which fails
to satisfy condition (i). 
Without loss of generality, take $\mathbf{R}_1$ to be this reach, with the
exclusive part $\mathbf{P}_1$, and the common part 
$\mathbf{Q}_1$.  
Our assumption means that there exists $z\in \mathbf{R}^n$ such that 
$z\neq 0$ and $z\in \left(\bigcap_{i\in\mathbf{P}_1}
  \mathcal{C}_i\right)\bigcap\left(\bigcap_{i\in \mathbf{Q}_1}
  \mathcal{C}_i\right)$.
Note that this implies $z\in\bigcap_{i\in\mathbf{P}_1}\Ker
  O_{C_i}$, $z\in\bigcap_{i\in\mathbf{Q}_1}\Ker O_{C_i}$,
  and $z\in\Ker\alpha_A^+(A)$.   

Consider the vector $y=b_1\otimes z\in\mathbf{R}^{nN}$. From
Lemma~\ref{lemma.undetect.H}, $y\in \bar{\mathcal{O}}$, and $y\neq 0$ since
$z\neq 0$. 
We now show that $y\in\prod_{i=1}^N\mathcal{C}_i$. According to the
discussion preceding the theorem, this vector can be partitioned as follows
$y=[y_1'\ldots y_N']'$, where $y_i=z$ for $i\in\mathbf{P}_1$,
$y_i=f^1_{1,i}z$ for $i\in\mathbf{Q}_1$, and $y_i=0$ for
$i\in\mathbf{V}\backslash \mathbf{R}_1$. Therefore, for every node
$i\in \mathbf{V}\backslash \mathbf{R_1}$, we have
$O_{C_i}y_i=0$. Also, for $i\in\mathbf{P}_1$, $O_{C_i}y_i=O_{C_i}z=0$
since
$z\in \bigcap_{i\in \mathbf{P}_1} \Ker O_{C_i}$.  
Similarly,  $O_{C_i}y_i=0$  for $i\in\mathbf{Q}_1$.
Since $z\in\Ker\alpha_A^+(A)$, then $y_i\in \Ker\alpha_A^+(A)$.
Thus, $y_i\in \mathcal{C}_i$  $\forall i$. 

We have shown that $y\in \bar{\mathcal{O}}\cap
\prod_{i=1}^N\mathcal{C}_i$. This leads to a contradiction with the
condition that $([\bar C',\bar H']',\bar A)$ is detectable;
see~(\ref{undetect.C.0}). This proves the statement of the theorem. 
\hfill$\Box$

\begin{theorem}\label{T4}
  Suppose the pair $(H,A)$ is observable. If for every reach $\mathbf{R}$ in
  $\mathbf{G}$, $\bigcap_{i\in\mathbf{R}} \mathcal{C}_i=\{0\}$, then 
the pair $([\bar C',\bar H']',\bar A)$ is detectable.
\end{theorem}

\emph{Proof: }
Suppose $([\bar C',\bar H']',\bar A)$ is not detectable, and therefore
there exists a nonzero vector $y\in \bar{\mathcal{O}}_H \cap
\prod_{i=1}^N\mathcal{C}_i$. From Lemma~\ref{lemma.undetect.H}, since the
pair $(H,A)$ is observable, 
then $\bar{\mathcal{O}}_H=\Ker\mathcal{L}\otimes \mathbf{R}^n$. Hence, the
vector $y$ can be represented as $y=b\otimes z$, where
$z\in \mathbf{R}^n$, and $b=\sum_{s=1}^k c_sb^s$; $c_1,\ldots, c_k$ are
scalar constants. Furthermore, using  Lemma~\ref{CV.lemma}, we have
$b_i= c_s$ if $i\in\mathbf{P}_s$, $b_i=c_sf^s_{s,i}$ if $i\in\mathbf{Q}_s$,
and $b_i=0$ otherwise.
Since $y\neq 0$, this implies that in the representation
$y=b\otimes z$, the vectors $z$, $b$ are nonzero. It further follows from
the condition $b\neq 0$ 
that for at least one $s\in\{1,\ldots,k\}$, $c_s\neq 0$ and
$c_sf^s_{s,i}\neq 0$ for all $i\in \mathbf{Q}_s$. Since the condition
$y\in\prod_{i=1}^N\mathcal{C}_i$ implies $c_sz\in \Ker\alpha_A^+(A)$,
$c_sO_{C_i}z=0$ for $i\in\mathbf{P}_s$ and $c_sf^s_{s,i}O_{C_i}z=0$ for
$i\in\mathbf{Q}_s$, this leads to the conclusion
that $z\in \cap_{i\in \mathbf{R}_s} \mathcal{C}_i$, which contradicts the
condition $\cap_{i\in \mathbf{R}_s} \mathcal{C}_i=\{0\}$. Hence $([\bar
C',\bar H']',\bar A)$ must be detectable.         
\hfill$\Box$

\begin{remark}
Since a digraph spanned by a tree is a reach, the result of
Corollary~\ref{U7.prop.3} can be seen as a special case of Theorem~\ref{T4}.
\end{remark}

\section{Example}\label{example}
In this section, we revisit Example 1 presented in~\cite{U6}. 
The state equation of the reference plant in that example is
6-dimensional and is governed by the $6\times 6$
state matrix
\begin{eqnarray} \nonumber
 &&A=\left[\begin{array}{rrrrrr}
    0.3775 &        0  &       0 &        0 &        0 &        0 \\
    0.2959 &   0.3510  &       0 &        0 &        0 &        0 \\
    1.4751 &   0.6232  &  1.0078 &        0 &        0 &        0 \\
    0.2340 &        0  &       0 &   0.5596 &        0 &        0 \\
         0 &        0  &       0 &   0.4437 &   1.1878 &  -0.0215 \\
         0 &        0  &       0 &        0 &   2.2023 &   1.0039
       \end{array}
     \right]. \label{A}
\end{eqnarray}
The plant is observed by the network consisting of six $H_\infty$
filters of the form (\ref{UP7.C.d}) connected in the topology of a directed
ring. The 1st
filter measures the 1st and the 2nd coordinates of the state vector,
the 2nd filter measures the 2nd and the 3rd coordinates, etc, with the last
filter taking measurements of the 6th and the 1st coordinates. In particular, 
$C_2=[\mathbf{0}_2~I_2~\mathbf{0}_{2\times 3}]$, 
$C_4=[\mathbf{0}_{2\times 3}~I_2~\mathbf{0}_2]$.

It can be directly verified that all eigenvalues of $A$ are in the right
half-plane, hence at every node of the network, the unobservable
modes of $A$ are not detectable. That is, $\mathcal{C}_i=\Ker
O_{C_i}$. Furthermore,  $\Ker O_{C_2}$ is spanned by the vectors
$d_4,d_5$, and $d_6$, while $\Ker O_{C_4}$ is spanned by $d_2,d_3$; here $d_i$
is the $i$th vector in the canonical orthogonal basis in
$\mathbf{R}^n$. Hence,
$\cap_{i=1}^6\mathcal{C}_i={0}$.  Also, $H=I_6$ in Example 1 of \cite{U6}. This
guarantees that $(H,A)$ is observable. Finally, the
6-node directed ring has a spanning tree. Thus, we have verified
all conditions of Corollary~\ref{U7.prop.3}. According to
Corollary~\ref{U7.prop.3}, the pair $([\bar C'~\bar H']', \bar A)$ in this
example is detectable, despite all the individual pairs $(C_i,A)$ having 
nontrivial undetectable subspaces.

To confirm this finding, the detectability of the matrix pair $([\bar
C',\bar H'],\bar A)$ was verified directly using Matlab. Also
in~\cite{U6}, a set of stabilizing output injection gains was found and the
stability of the system (\ref{e.0}) was verified directly, thus confirming
our conclusion based on Corollary~\ref{U7.prop.3}.

It follows from Corollary~\ref{U7.prop.3} that the detectability of $([\bar
C',\bar H'],\bar A)$ will be
preserved even if the filters transmit, e.g., only the third and fifth
coordinates of their respective estimate vectors, that is, if instead of
$H=I_6$, we take 
$
H=\left[\begin{array}{cccccc} 0 & 0 & 1 & 0 & 0 & 0 \\
                                   0 & 0 & 0 & 0 & 1 & 0 
                                 \end{array}\right].
$
With this $H$, $(H,A)$ is observable, and
Corollary~\ref{U7.prop.3} is still applicable. This creates a potential
for reducing the amount of information transmitted by the nodes,
since only two coordinates of the estimate vectors $\hat x_j$ need to be
transmitted instead of all six coordinates.      
However, if the filters transmit the 2nd and the 5th
coordinates of $\hat x_j$, the pair $(H,A)$ will not be observable and
the system cannot be guaranteed to be 
detectable. In fact, one can check directly that the corresponding pair
$([\bar C',\bar H'],\bar A)$ is not detectable. Therefore, the
distributed filter of the form (\ref{UP7.C.d}) cannot be constructed in
this case.

\section{Conclusions}\label{Concl}
The paper presents necessary and sufficient conditions for
detectability of a linear plant via a network of state estimators. 
We showed that the detectability of the entire system can be ascertained
from the detectability properties of the filters' pairs $(C_i,A)$, along
with the matching properties of interconnections. Our results complement the existing results on distributed consensus-based
estimation by elucidating the relationship between the network topology and
detectability/observability properties of the plant and filters. Future
work will investigate a similar relationship between the network topology and
controllability of multi-agent systems. 


\newcommand{\noopsort}[1]{} \newcommand{\printfirst}[2]{#1}
  \newcommand{\singleletter}[1]{#1} \newcommand{\switchargs}[2]{#2#1}

\end{document}